# Current-Dependent Periodicities of Si(553)-Au

S. Polei[1], P.C. Snijders[2*], K.-H. Meiwes Broer[1] and I. Barke[1**]

[1]Universität Rostock, Institut für Physik, Rostock, Germany,

[2]Materials Science & Technology Division, Oak Ridge National Laboratory, Oak Ridge, Tennessee 37831, USA, and Department of Physics and Astronomy, The University of Tennessee, Knoxville, Tennessee 37996, USA

**Abstract**

We investigate quasi one-dimensional atomic chains on Si(553)-Au with a scanning tunneling microscope (STM). The observed periodicity at the Si step edge can be altered by the STM and depends on the magnitude of the tunneling current. In a recent report this reversible structural transition was attributed to transient doping with a characteristic time scale of a few milliseconds [1]. Here we explore the evolution of the STM topography as a function of the magnitude of the tunneling current for a wide temperature range. Based on a decomposition of topographic line profiles and a detailed Fourier analysis we conclude that all observed current-dependent STM topographies can be explained by a time-averaged linear combination of two fluctuating step-edge structures. These data also reveal the precise relative alignment of the characteristic STM features for both phases along the step edges. A simple diagram is developed, presenting the relative contribution of these phases to the STM topography as a function of tunneling current and temperature. Time- and current-dependent measurements of fluctuations in the tunneling current reveal two different transition regimes that are related to two specific current injection locations within the surface unit cell. A method based on spatially resolved $I(z)$ curves is presented that enables a quantitative analysis of contributing phases.

# I. Introduction

The capability to manipulate surfaces on the atomic level with a scanning tunneling microscope (STM) has led to new possibilities to tailor their structural and electronic properties. This enabled the discovery of exciting phenomena such as the quantum mirage effect [2], or, recently, the construction of a single atom transistor [3] and spin-logic circuitry [4]. Techniques for surface modification include tip induced rearrangement [5], diffusion [6,7] or desorption [8,9,10] of atoms and molecules. Moreover, the manipulation of even larger structures such as nanometer-sized islands or carbon nanotubes can be accomplished via localized injection of excess charge using a tunneling current [11,12,13].

Recently we have reported on the effect of transient electron doping on the Si(553)-Au surface structure. In this technique, electrons tunnel from the tip into the sample and have a finite probability to briefly stay in the surface electronic system before eventually draining to the bulk. This transient electron doping destabilizes the original 1×3 ordered ground state of the surface structure, which reorganizes into a 1×2 ordered phase. When the injected charge drains into the bulk, the system relaxes to its 1×3 ground state, where it remains until the next doping process occurs. By varying the tunneling current, the transition rate from the 1×3 to the 1×2 phase could be controlled. Because the magnitude of the tunneling current at constant tip height is affected by the transient changes of the surface structure, recording time-resolved current traces enabled the extraction of the corresponding lifetimes $\tau_0$ and $\tau_1$ of 1×3 ground- and 1×2 excited state, respectively [1].

Here the focus is on an alternative analysis based on time-averaged features, i.e. in the regime where the measurement rate is slower than the timescale of the transient fluctuations of the surface structure. This approach allows us to analyze the topographic changes due to the phase transition yielding important information for the development of a structural model of the excited state. Based on a systematic study of current and temperature dependent topographic periodicities, a qualitative phase diagram is established that visualizes the evolution of the one-dimensional surface structure. Spatially resolved *I(z)* curves are utilized to reveal intra-unit cell differences in the doping efficiency. These differences are attributed to the spatial distribution of the amplitude of the wave function associated with the state into which the doping charge is injected. Finally, a method is presented that quantitatively describes the current dependence of time-averaged topographic STM data, and allows for an independent extraction of typical currents necessary to excite the system.

## II. Methods

Experiments were performed in a commercial low temperature STM system equipped with a separate preparation chamber. In both chambers the base pressure was < $10^{-10}$ mbar. The STM was operated at temperatures between 7 K and 78 K. Positive tunneling biases correspond to electrons traveling from tip to sample. The Si(553) substrate (0.01-0.03 Ohm-cm, *p*-type) was first degassed at 650 °C for a few hours and subsequently flashed several times to 1280 °C. The optimum Au coverage to obtain a well-ordered Si(553)-Au surface is ~0.5 ML [14]. The best surface quality is obtained by depositing slightly more than 0.5 ML Au on the substrate at 650 °C, followed by a brief post anneal at 1060 °C to desorb excess gold. This resulted in a well-defined Si(553)-Au reconstruction with low defect density.

Current-distance (*I(z)*) curves recorded during a linear *z* ramp were numerically inverted to yield *z(I)*. Measurements were performed with Cr and W tips that were electrochemically etched in HCl and NaOH solutions, respectively, followed by in vacuo conditioning through heating and Ar ion self-sputtering. The results presented in the following were consistent for experiments using Cr and W tips.

## III. Results
### A. Current-dependent corrugation

In Fig.1a) the STM topography of the Si(553)-Au surface at 60 K is shown. Bright vertical stripes with a perpendicular spacing of ~1.5 nm represent the Si step edges [15]. For sufficiently low tunneling currents (2 pA) these step edges exhibit bright protrusions that are equally spaced along the chains by $3 \cdot a_{Si}$ (~ 1.15 nm), establishing a 1×3 periodicity characteristic of the low-temperature phase [16-19].

The same sample area imaged at higher current (2 nA) is presented in Fig.1b). A clear change in the periodicity to a 1×2 structure is observed, showing that the structure is controlled by the tunneling current.

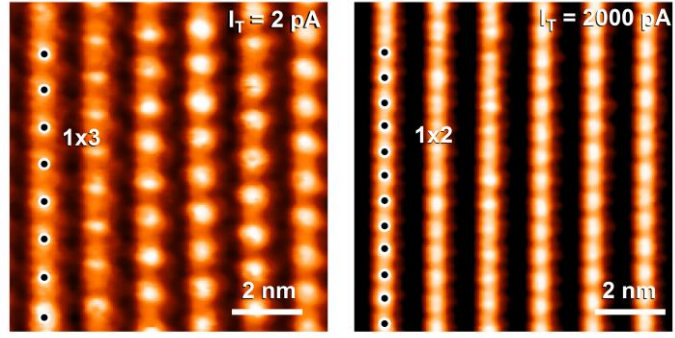

FIG.1: STM topography ($U_{gap}$ = 1 V) of Si(553)Au taken at 60 K and different tunneling currents. Left: at $I$ = 2 pA bright protrusions on the step edge chains form a 1×3 periodicity. Right: same sample area at $I$ = 2 nA. On the step edges a 1×2 periodicity is observed.

In a previous publication we have shown that this transition is accompanied by a fast switching between the 1×3 and 1×2 structures [1]. Recording topographic STM images $z(x,y)$ at constant current $I$ at a pixel imaging rate slower than the typical frequency of these fluctuations is then expected to correspond to a simple linear combination of the 1×3 ($z_{1x3}(x,y)$) and 1×2 ($z_{1x2}(x,y)$) phases:

$$z(x,y,I) = \alpha(I) \cdot z_{1\times 3}(x,y) + (1-\alpha(I)) \cdot z_{1\times 2}(x,y) + \beta(I), \qquad (eq.1)$$

where $\alpha(I) = \dfrac{\tau_0}{\tau_0 + \tau_1}$ is the (current dependent) fraction of time that the surface displays 1×3 order and $\beta(I)$ is a current dependent offset in z due to the reduced tip-sample distance when increasing the tunneling current. We assume here that $z_{1x3}(x,y)$ and $z_{1x2}(x,y)$ are independent of $I$, justified *a posteriori* by the good agreement in Fig.2e), see below.

High resolution images of a single Si step edge at 60 K are presented in Fig.2a)-c) together with the corresponding line profiles in d)-f) (solid black curves). The 1×3 structure at $I$ = 3 pA consists of a sequence of bright protrusions with alternating apparent height. For a slightly higher current of $I$ ~ 10 pA (Fig.2b)) a 1×6 periodicity appears, the corrugation of which is dominated by the underlying 1×1 structure. At $I$ = 2.1 nA the 1×2 structure with a characteristic double-peak feature is fully developed (Fig.2c)). Additionally the Au chain below the step edge becomes visible in b) and c) in form of a sequence of less intense 1×2 ordered features. This 1×2 period on the Au chains is consistent with observations in previous reports [16,18,20].

The red dotted curve in Fig.2e) shows the result of a fit according to Eq.1, yielding $\alpha = 0.55$. In this fit a small constant shift along x for each line profile is included to account for limited thermal drift between measurements. The linear combination is in good agreement with the measured 1×6 profile, confirming that *the 1×6 periodicity observed in Fig.2b) and e) does not represent an independent structural phase but is rather formed by a simple superposition of the 1×2 and 1×3 periodicities* that fluctuate faster than our STM data acquisition rate, resulting in a time-averaged 1×6 corrugation in the STM images.

Note that theoretical investigations have shown that the ground state for Si(553)-Au has an antiferromagnetic arrangement of magnetic moments on every third silicon atom on the Si step edge [15]. The authors predict that this antiferromagnetic spin polarization should appear as a 1×6 periodicity in spin-resolved STM experiments. Indeed some of the data presented here have been taken using Cr tips that have been shown to be suited for spin resolved measurements [21]. However, our observations were reproduced using a conventional W tip incapable of spin-dependent resolution. Hence the 1×6 structure reported in this work is evidently not directly caused by antiferromagnetic order, but rather by a simple linear superposition of 1×3 and 1×2 structural order regardless of a possible spin structure of either of the two phases. The observation of a 1×6 periodicity by itself without a detailed current dependent analysis is therefore not sufficient to discriminate between these two different physical mechanisms.

At 60 K, the dominant peak of the 1×3 line profile (Fig.2d), black solid curve) can be observed to be constituted of two smaller features. This can be attributed to a slight admixture of the 1×2 structure even at this low current (compare Fig.1 in Ref. [1]). At higher temperatures this splitting is absent and the dominant peak is narrower, as shown in Fig.2d) (dashed curve) for $T = 78$ K and $I = 10$ pA. An accurate determination of $\alpha$ using equation (1) would require the use of pure 1×2 ($\alpha = 0$) and pure 1×3 ($\alpha = 1$) topographies of the same sample area and at the same temperature. As we will show below this is a non-trivial requirement due to the wide current range where structural fluctuations due to transient doping can be observed. The optimal temperature to perform this analysis turns out to be 60 K, although even at the lowest currents used a small admixture of 1×2 still remains in the predominantly 1×3 periodicity. This results in a systematically overestimated value for $\alpha$ at that temperature [22].

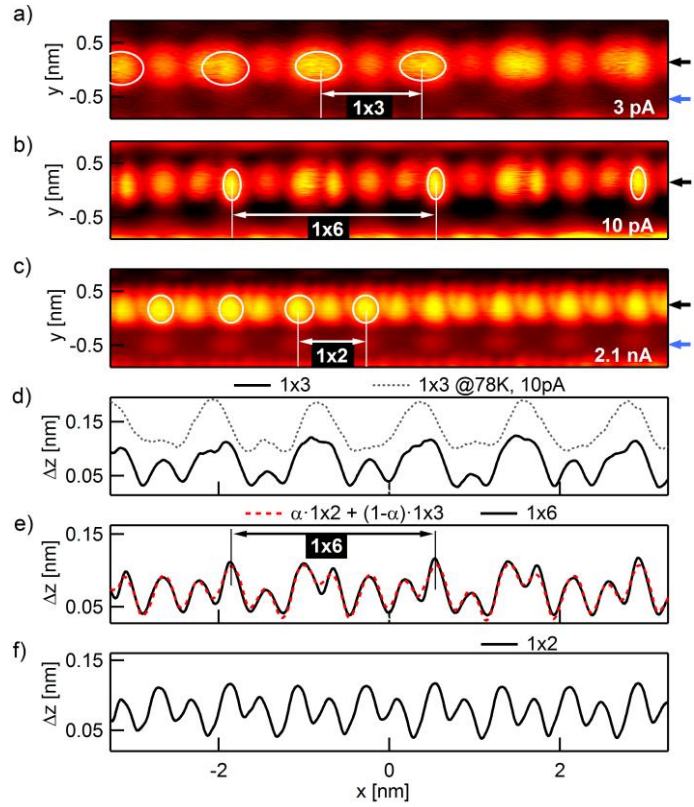

FIG.2: STM topography ($U_{gap}$ = +1 V, $T$ = 60 K) of a single Si-step edge. a) 3 pA: 1×3 periodicity, b) 10 pA: 1×6 periodicity, c) 2.1 nA: 1×2 periodicity. d) - f): corresponding line profiles at locations indicated by the black arrows on the right side of the topography images. The red dashed curve in e) corresponds to a linear combination of the 1×3 and 1×2 line profiles in d) and f) with $\alpha$ = 0.55 (see Eq.1). This shows that the 1×6 structure is a (time-averaged) linear combination of the two other phases. In d) a line profile at 78 K and $I$ = 10 pA is added (dashed gray line) to illustrate that at 60 K and 3 pA the topography already contains an admixture of the 1×2 phase. The blue arrows refer to data in Fig.3.

## B. Alignment of 1×3 vs. 1×2 at the step edge

In order to understand the properties of the 1×2 phase, and the reversible 1×3-to-1×2 transition, it is imperative to identify the structure of the 1×2 phase. However, the transiently doped nature of the 1×2 phase makes this a difficult proposition. As a starting point, we therefore analyze the relative location of characteristic 1×2 and 1×3 features. From the alignment of the 1×2 and 1×3 contributing phases in their superposition that results in the 1×6 corrugation (see Fig.2), we infer that the maxima of the 1×2 phase are offset from the dominant maxima in the 1×3 phase. To accurately determine this offset, one possible approach is to find a fixed internal reference, present in both the 1×3 and the 1×2 topographic images. Although very faint in low current images, the 1×2 period of the Au chain located on the terrace adjacent to the step edge [15] is well suited for that purpose. Its structure is not directly affected by the phase transition (compare [1]). Line profiles of the step edge and of the adjacent Au chain (taken along the directions indicated by the black and blue arrow in Fig.2c) for the high-current 1×2 phase are extracted from a single image and compared in Fig.3a). It is evident that the 1×2 periods on the step edge and on the Au chain are aligned in phase.

In a second step low current line profiles of the step edge and the Au chain, taken along the directions indicated by the black and blue arrows in Fig2a), are compared in Fig.3b). Although barely visible in the topography, the 1×2 period of the Au chain can be clearly identified in the line profile. Consistent with our preliminary conclusion above, the maxima of the Au chain do not coincide with the dominant $b_i$ 1×3 maxima of the step edge at any location. Analyzing the minimum distance between maxima of the step edge and of the Au chain reference reveals an average displacement of

$$b_{mean} = \frac{1}{n}\sum_{i=1}^{n}|b_i| = 0.17 nm \pm 0.02 nm \quad \text{(see Fig.3).}$$

This value is consistent with the lateral offset of $b_{model}$ = 0.19 nm between Si atoms directly at a [1-10] step edge and Si atoms one row behind the step edge on the same terrace (see Fig.3c)), and [15]).

This conclusion is supported by an independent and complementary analysis based on a measured 1×6 line profiles (Fig.2e)). Here we utilize the fact that the 1×6 line profile contains information on both structures simultaneously. The best fit shown in Fig.2e) not only yields the fraction of each component but also their relative alignment by comparing the location of the maxima of the constituent 1×2 and 1×3 line profiles. We obtain $b_{mean}$ = (0.18 ± 0.03) nm, in good agreement to the value obtained from a fixed internal reference.

These results indicate that the high current 1×2 step edge corrugation is located one atomic row behind the low current 1×3 corrugation, see Fig. 3c). This also implies that the maxima of the 1×2 phase should be located ≈ 0.1 nm away from those of the 1×3 structure in the direction perpendicular to the chains. While our data do suggest consistence with such an offset (not shown) a quantitative verification is far more difficult due to the larger effect of drift along the y-direction (slow scanning direction). This uncertainty is further compounded by the shape of the step edge features in the STM images that is generally not circularly symmetric, which decreases the precision in a quantitative measurement perpendicular to the step edge chains.

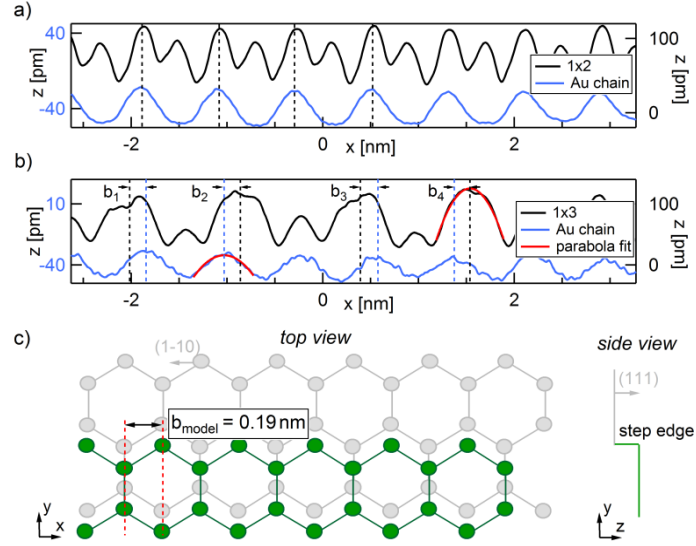

*FIG.3:* a) and b): line profiles of the step edge (black) and of the Au chain (blue) extracted from the STM images shown in Fig.2c) and Fig.2a), respectively. The line profiles were taken along the directions indicated by the black and blue arrows in Fig.2a) and Fig.2c), respectively. Red lines represent parabolic fits used to determine the positions of the peak maxima. In a) the line profiles of the step edge and the Au chain both exhibit a $1\times2$ periodicity. The maxima are aligned in phase, whereas in b) a clear displacement between dominant maxima of step edge and the Au chain is evident. An average offset of $b_{mean} = 0.17$ nm is found. c): Schematic structure of the underlying Si lattice of the step edge on Si(553) to illustrate the distance between atoms directly at the step edge and atoms one row behind: $b_{model} = 0.19$ nm.

## C. Temperature and current dependence of the transient phase transition

The strong temperature dependence of the $1\times3$ to $1\times2$ structural transition results in the $1\times3$ not being observable at very low temperatures [1]. Moreover, the appearance of the transition as a function of tunneling current is continuous (see Fig. 2b),c)) and there is no obvious well-defined threshold current for the transition. In the following the temperature and current dependence of the transition is systematically analyzed. Topographic STM data were recorded for different tunneling currents in a temperature range between 7 K and 78 K. Despite the lack of a clearly defined threshold current, we can attempt to describe the progress of the transition using the Fourier coefficients for the two different periodicities extracted from line profiles, see Fig.2d)-f) (see also [23] for a related analysis): we introduce *r* being the difference of the Fourier coefficients $a_{1\times2}$ and $a_{1\times3}$ normalized by their sum, similar to the definition of the effective polarization *P* of a spin-polarized tunnel junction [24]:

$$r = \frac{a_{1\times2} - a_{1\times3}}{a_{1\times2} + a_{1\times3}} \qquad \text{(Eq.2)}$$

Note that although *r* is well suited for comparing the transition at different temperatures for a given current, it does not represent the correct ratio of the 1×3 vs. the 1×2 occurrence probabilities as a function of current. This is caused by the normalization and most importantly by the fact that $a_{1\times3}^{max} \neq a_{1\times2}^{max}$ (i.e. the corrugation amplitudes of the pure phases are not equal in magnitude) and $a_{1\times2}^{min} \neq a_{1\times3}^{min} \neq 0$. However, at a given current (temperature), the dominance of 1×3 and 1×2 phases in the topographic STM images at different temperatures (currents) can be well compared by evaluating *r*, see Fig.4. The experimentally accessible range of *I*, indicated by dotted vertical lines, is limited to currents between 2 pA and 3 nA, due to preamplifier noise at low currents, and frequent tip changes at high currents.

For 60 K the current dependence is displayed for chain segments at two different locations on the surface (solid red circles with dark and light gray stroke colors respectively) to illustrate the typical spread of *r* for a given temperature. Up to ~ 5 pA the 1×3 structure is found to be stable resulting in a small and almost constant *r* as a function of the tunneling current. Upon further current increase the 1×2 contribution increases monotonically. This behavior continues up to the highest applied current of ~ 3 nA. The corresponding large positive *r* reflects the dominating 1×2 periodicity for that current.

For 78 K and 50 K this trend is shifted horizontally. At 78 K higher currents are needed to induce the same degree of 1×2 admixture as compared to 60 K (solid orange circles in Fig.4). The opposite behavior is observed at 50 K (solid dark red circles), where small currents already result in a substantial 1×2 contribution. We note that the exact current for a specific value of *r* and *T* (i.e. the horizontal location in the diagram of Fig.4) is found to vary slightly between experiments. We attribute this to varying tip conditions or different charge decay rates. The latter likely depends on the bulk doping profile in vicinity of the surface (compare Ref. [1]) which can be affected by the sample preparation process [25,26].

By comparing *r* at 78 K, 60 K, and 50 K for similar currents an interesting temperature dependence is evident: the 1×2 phase as observed in topographic STM images can be most easily triggered at low temperatures, in contrast to the typical case of STM induced structural changes where low temperatures usually result in a rigid system that is more difficult to manipulate [6,27,28]. Joule heating due to energy dissipation from the tunneling current as a driving force for the structural transition can thus be excluded.

Measurements at even lower temperatures (13 K and 7 K, solid black circles in Fig.4) reveal that the excited 1×2 structure is fully established, almost independent of the applied current. This shows that for lower temperatures even small currents are sufficient to induce the structural transition from 1×3 to 1×2. Extrapolating the current necessary for $r = 0$ (i.e. equal 1×3 and 1×2 amplitude) from higher temperatures down to 7 K results in numbers of the order of some fA or less which is far below current experimental detection limits. An interesting implication is that the ground state 1×3 structure of the Si(553)-Au surface is in practice not accessible in a low-temperature STM experiment. Instead, higher temperatures are required to minimize the effect of transient electron doping. We note that careful atomic force measurements may be a possibility to circumvent this fact, if the tunneling current due to residual potential differences between tip and sample can be kept sufficiently low.

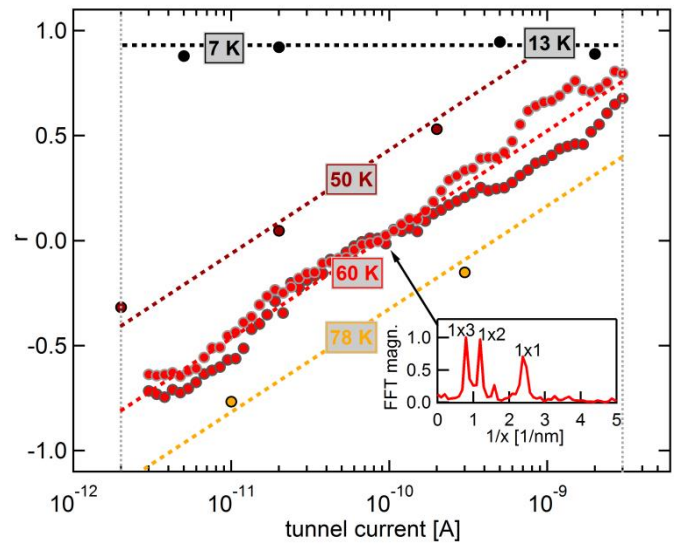

FIG.4: Simple phase diagram of the 1×3 ↔ 1×2 transition: normalized ratio $r$ of the Fourier coefficients $a_{1\times 2}$ and $a_{1\times 3}$ for different tunneling currents and temperatures (circles). Larger values of $r$ (i.e. the periodicity is increasingly dominated by 1×2 order) are found for increasing current or decreasing temperature. Colored dashed lines are guides to the eye. The experimental current limits in our experiments are indicated by vertical dotted lines. A typical Fourier spectrum for 60 K and ~ 100 pA is displayed in the inset, where peaks corresponding to different periodicities are indicated. For 7 K and 13 K $r$ is found to be constant because for low temperatures the system is always excited to the 1×2 phase, even at the lowest currents.

**D. Spatial dependence of the transient doping efficiency**

So far we neglected the possibility that the efficiency of the doping process is inhomogeneously distributed within the 1×6 unit cell which encompasses both phases. However, the transient doping efficiency, and thus the probability for a 1×3 to 1×2 transition to occur, is expected to be dependent on the specifics of the electron state where the doping charge is injected in, such as its spatial distribution, orbital character and energy. Since the amplitude of the wave function associated with that state is not constant across the unit cell, this should be visible as a varying 1×3 ↔ 1×2 fluctuation rate depending on the location within the unit cell. A convenient way to record these fluctuations is to measure the tunneling current at constant bias while slowly varying the tip-sample distance. Fig.5a) shows such *I(z)* curves measured on top of ($x_1$, black) and in between ($x_2$, red) bright 1×3 protrusions. At $x_1$ fluctuations are clearly visible for a mean current of $\bar{I} \sim 10$ pA, originating from the telegraph signal reported in [1]. To efficiently map the occurrence of fluctuations, we calculate the relative standard deviation $\sigma_{rel}(\bar{I}) = \sigma(\bar{I})/\bar{I}$ with $\sigma(\bar{I}) = \sqrt{\frac{1}{N-1} \sum (I - \bar{I})^2}$, where *N* is the number of data points within a moving window of width $\Delta z = 10$ pm. In Fig.5b) $\sigma_{rel}(\bar{I})$ is plotted as a function of the mean tunneling current for positions $x_1$ and $x_2$. For better statistics, the standard deviations of several *I(z)* curves at equivalent locations were averaged. At $x_1$ the maximum is located close to the lowest measurable current (black curve), whereas at $x_2$ it is found at $\bar{I} \sim 300$ pA. Since the current-induced excitation is a statistical process, an exact threshold current for the transition does not exist. Nevertheless we can use the maxima of the standard deviation to define typical transition currents, as they identify the current where the fluctuations are most pronounced. We infer that the structural transition is more easily (i.e. at lower currents) triggered at the bright protrusions of the 1×3 reconstruction than in between these protrusions. Hence, the transition current is strongly dependent on the location of the excitation, confirming a dependence of the doping efficiency on the spatial distribution of the specific surface state, and aiding in a possible future identification of the electronic state that is involved in the transient doping process here.

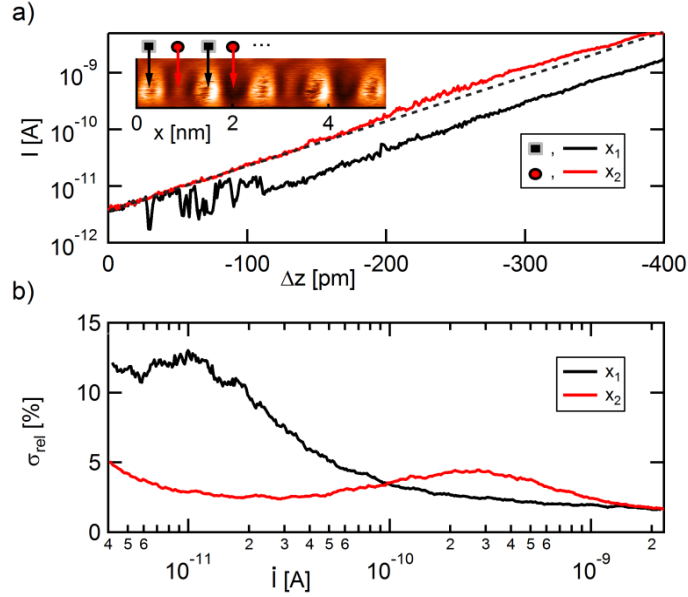

FIG.5: a): Typical $I(z)$ curves ($U_{gap}$ = 1.3 V, $T$ = 57 K) measured over a bright 1×3 protrusion (black; position $x_1$) and in between (red; position $x_2$), respectively. The dotted line is a guide to the eye representing a purely exponential behavior. Inset: topography ($U_{gap}$ = 1.3 V, $I$ = 10 pA, $T$ = 60 K) of a single Si step-edge showing a 1×3 periodicity and the locations of $x_1$ and $x_2$. Fluctuations on top of bright 1×3 protrusions can be seen as dark horizontal stripes. b): The relative standard deviation of the tunneling current as a function of the mean current measured on positions $x_1$ and $x_2$. The curves represent an average over ~ 200 measurements. The peaks in the relative standard deviation reveal the typical currents where the 1×3 ↔ 1×2 structural fluctuations occur and show that at these currents differ for the two locations $x_1$ and $x_2$.

In the fitting of the time-averaged line profiles (Fig.2e)) we implicitly assumed that the imaged step edge exhibits a spatially uniform phase transition. Due to the location-dependent magnitude of the transition current revealed in Fig.5b) this is only approximately correct. Hence small local variations in how well the linear combination of $z_{1\times 3}(x,y)$ and $z_{1\times 2}(x,y)$ (see eq.1) describes the observed corrugation should be expected due to the differences in excitation probability as observed in Fig.5b). The prefactor $\alpha$ in eq.1 is in truth a function of position $x$ within a 1×6 unit cell which is neglected in our analysis.

### E. Quantitative contribution of each phase to the measured corrugation

In the following we present a method, which yields a quantity that is directly proportional to the occurrence probability of one of the two phases at any current, even if both pure phases (i.e. without admixture of the respective other phase) are not accessible. This enables a quantitative comparison of the contribution of a phase to topographies measured with different currents. Neither the standard deviation (section D) data nor the Fourier analysis in Fig.4 (section C) are suited for this purpose. This is due to the amplitudes of the fluctuations being dependent on the local differences of the corrugation amplitude of 1×3 and 1×2, see section C. Hence, a more sophisticated analysis is necessary, that does not rely on the availability of the pure phases, and that takes into account the spatial dependence illustrated in Fig.5. For this purpose tip-sample displacement measurements *I(z)* were inverted to obtain *z(I)*. This approach yields information on the change of the apparent height as a function of the current at a particular location.

In Fig.6 averaged *z(I)* curves taken on two different locations A (green) and B (black) are presented. The logarithmic scale of the abscissa results in an almost linear evolution with a systematic divergence between the curves that is largest for low tunneling currents. This directly reflects the fact that positions A and B are inequivalent for the low-current 1×3 structure (for a detailed description, see Appendix A).

The difference between curves A and B ("A-B", gray dots in Fig.6) monotonically decreases as a function of *I* and approaches zero for large currents. As explained in Appendix A, this difference is a measure for the fraction of 1×3 contained in the observed current-dependent structure. A kink is visible at ≈ 30 pA, separating the curve into two sections, one with a rapid structural change at low currents and one with a less pronounced change for *I* > 30 pA. Since the tunneling current represents the excitation mechanism an exponential dependence $\Delta z(I) = \Delta z_0 \exp(-I/\kappa)$ is expected from the transient doping scenario, in close analogy to the optical excitation of atomic or molecular species [29,30]. The decay parameter $\kappa$ is a measure for the typical current necessary to excite the system, and $\Delta z_0$ is the topographic difference A-B at zero current. Since two different regions of topographic fluctuations are identified (see Fig.5), a double exponential function $\Delta z(I) = \Delta z_0^{(a)} \exp(-I/\kappa_a) + \Delta z_0^{(b)} \exp(-I/\kappa_b)$ is fitted to the data (blue curve in Fig.6). The good agreement of the data with this model independently confirms the current dependence of the time constants $\tau_0$ and $\tau_1$ in Ref. [1] using a fundamentally different method. The fit yields $\kappa_a$ ~ 10 pA and $\kappa_b$ ~ 250 pA which are in excellent agreement with the locations of the maxima in the relative standard deviation for the fluctuation analysis of the tunneling current in Fig.5b). As mentioned above, the double exponential curve A-B is directly proportional to the fraction of the $1\times3$ phase as a function of the tunneling current. Note that knowledge of the pure $1\times2$ phase is not necessary to deduce this proportionality, i.e., the method is still applicable if only part of the transition is tracked experimentally. If in addition the pure $1\times3$ phase is available, the absolute $1\times3$ contribution $frac_{1x3}(I)$ can be obtained by normalization to the corresponding $\Delta z$. This is illustrated in Fig.6 by a separate axis (red) for the A-B data. The scaling error (red dashed lines), which consists of a constant scaling factor for the entire curve, is a result from a slight $1\times2$ admixture in the step edge topography image even at 3 pA (see also inset of Fig.6 and section A). Details regarding the error estimation are provided in Appendix B. The fact that in Fig.6 the fraction of the $1\times3$ phase is quantitatively described for the entire current range allows us to connect this analysis to those of the previous sections by considering one particular point of the $frac_{1x3}(I)$ curve: at a current of about 15 pA both phases occur with roughly equal probability. This is in good agreement with the line profile analysis in section A (55% $1\times3$ at 10 pA), the typical current (10 pA at $x_1$) obtained from the standard deviation in section D, and the parameter $\kappa_a$ = 10 pA from the double-exponential fit in this section. Note that topography measurements and I(z) curves (i.e. without lateral tip movement) yield the same current value for equal probability of both phases. Hence the scan speed has no influence on the phase transition for the scan speeds used here (<40 nm/s).

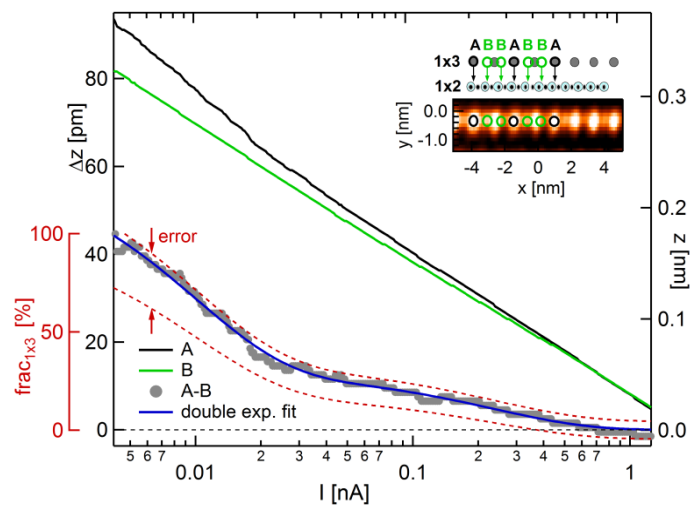

FIG.6: Tip displacement *z(I)* as function of tunneling current for positions A and B (black and green curve) as indicated in the bottom of the inset. The data are calculated from measured *I(z)* curves and represent an average over equivalent locations along the chain. The locations are chosen such that all contributions from the 1×2 phase cancel out in the difference A-B (gray dots, see Appendix A for a detailed description). The values of A-B are proportional to the fraction of the 1×3 contribution. A double exponential fit of A-B matches the data, reflecting the saturation behavior of the phase transition induced by transient doping at increasing current. Two exponential functions are necessary, one for each location with pronounced fluctuations (cf. Fig.4). Within the scaling error indicated by red dashed lines the absolute amount of 1×3 can be read from the red axis (description see text). Inset (top): scheme of inequivalent locations (A and B) relative to 1×3. Inset (bottom): 1×3 topography measured at 3 pA and simultaneously recorded with the *I(z)* curves. Red and black circles indicate positions A and B, respectively.

## IV. Discussion and conclusion

In conclusion we have presented time-averaged topographic STM images of the Si(553)-Au surface that show a transition of the periodicities at the Si step-edge. The measured structural periodicity changes from 1×3 to an apparent 1×6, and finally to 1×2 with increasing tunneling current or decreasing temperature. From an analysis of line profiles of the topographic corrugation (Fig.2) we have shown that the gradual appearance of the transition can be described by a linear superposition of the two fluctuating contributing phases. One consequence is that a 1×6 structure is observed that should not be confused with the antiferromagnetic spin-polarized 1×6 structure predicted earlier [15]. By analyzing the relative alignment between both phases we conclude that the STM features observed for the 1×2 structure originate from locations that correspond to one silicon row behind the step edge, toward the upper terrace. This finding is an important prerequisite for the development of an atomic model of the excited phase. Our temperature dependent studies of the observed periodicities enable the construction of a qualitative diagram (Fig.4) of contributing phases as a function of tunneling current and temperature. This phase diagram reveals that the transition to the current-induced 1×2 state systematically shifts to smaller currents when reducing the temperature. This rather unusual temperature dependence is fully consistent with the previously proposed scenario of transient doping by the STM tunneling current [1]. The analysis of fluctuations in the *I(z)* data at different locations reveals that the typical transition current is significantly higher between two bright 1×3 protrusions than on top of them (Fig.5), reflecting a spatially varying excitation probability. Finally, we present a method based on tip-sample displacement measurements that yields a quantitative and position-selective analysis of the phase transition. An exponential saturation of the excited state as function of the tunneling current is obtained (Fig.6) and two different transition currents are found, in perfect agreement with the fluctuation analysis.

The findings reported here in principle bear relevance to any STM study of systems with a low-dimensional electronic structure. An important finding is that low temperature STM experiments do not necessarily relate to the ground state of the system; instead higher temperatures may be needed to access the ground state structure. Indeed, STM or tunneling current induced changes similar to those reported here have been implicated in for example the (low temperature) surface structures on Sn/Ge(111) and Si(100) [31,32]. However the identified mechanism of STM induced structural changes on the Si(553)-Au surface is fundamentally different from those reported for the Sn/Ge and Si(100), as pointed out in [1]. The analysis presented here provides a relation between time averaged and dynamic properties for the case of the doping-induced transition on Si(553)Au. It is conceivable that this connection is of relevance for related systems as well.

Finally we would like to note that one-dimensional systems are notoriously hard to dope as the perturbation imposed by the dopant atoms onto the structure effectively cuts the one-dimensional chains into finite sections [33]. The ability to precisely control a dynamic phase transition via transient doping as demonstrated here, may open new routes for systematic manipulation of low-dimensional electronic systems by accessing parts of a doping-dependent phase diagram that otherwise remain hidden.

**ACKNOWLEDGMENTS**

Funding by the federal state Mecklenburg-Vorpommern within the project *Nano4Hydrogen* and by the Federal Ministry of Education and Research (BMBF) within the project *Light2Hydrogen* is gratefully acknowledged (S. P.). P. C. S. acknowledges support by the U.S. DOE Office of Basic Energy Sciences, Materials Sciences and Engineering Division, through the Oak Ridge National Laboratory.

V. Appendix A:

The analysis presented in Fig.6 is based on a particular selection of spatial locations that enable extraction of the occurrence of a single structure, either $1\times3$ or $1\times2$, from the measured composite curves. Here we only extract the low-current $1\times3$ structure because it shows a higher corrugation compared to the 1×2 structure, thus yielding a better signal-to-noise ratio. In order to cancel out the 1×2 contribution, all tip displacement curves are extracted on a 1×2 grid along the chain, such that all curves contain the same 1×2 contribution. Any differences in individual curves are then not originating from the 1×2 structure:

Let $z^i(x) = z^i(T_i \cdot n + x)$ (with $n \in \mathbb{N}$) be the contribution of the periodic structure $i$ (either 1×2 or 1×3) with period length $T_i$ to the measured STM corrugation. The total topographic corrugation is then

$$z^{1\times6}(x) = z^{1\times3}(x) + z^{1\times2}(x).$$

Let $A = z^{1\times6}(T_{1\times6} \cdot n)$ and $B = z^{1\times6}(T_{1\times6} \cdot n + T_{1\times2})$ be the topographic data at the locations A and B, respectively (see Fig.6). The difference $A - B$ then yields

$$A - B = z^{1\times3}(T_{1\times6} \cdot n) - z^{1\times3}(T_{1\times6} \cdot n + T_{1\times2}),$$

which evidently does only contain information on the 1×3 phase.

The lateral offset (i.e. the starting point of the 1×2 grid) is chosen to maximize 1×3 contrast of two subsequent curves (see inset of Fig.6a)). Due to the existence of two inequivalent 1×2 locations relative to the 1×3 structure two types of curves exist, one taken on a bright 1×3 location (type "A", black circles) and the other with significantly lower z (type "B", red circles). All locations are determined using the simultaneously recorded topography at $U_{gap}$ = 1.3 V and $I$ = 3 pA where the 1×3 structure is observed (see inset of Fig.6a)).

This scheme thus allows us to quantitatively extract the contribution of one of the two phases to the *I(z)* curves despite the fact that all curves contain unknown contributions from both phases.

V Appendix B:

In Fig.6 two red dotted lines are added to the $\Delta z$ data, representing a worst case estimate of the scaling error (see outermost left axis). To estimate this error, the following scheme has been used. From the line profile at 3 pA (see Fig.2e)) a variable fraction of a (*x*-drift corrected) 1×2 line profile (see Fig.2f)) is subtracted. The fraction *c* is chosen to minimize the remaining 1×2 component in the Fourier spectrum, yielding *c* = 0.3. This means that a maximum 1×2 fraction of 30% is contained in the line profile at 3 pA. This value is an upper limit since for the analysis it is assumed that the topography of the pure 1×3 ground state does not contain any 1×2 Fourier components that match those of the measured 1×2 line profile (this includes Fourier components induced by experimental noise). An additional offset of ± 2 pm is added to the error of the $frac_{1x3}(I)$ curve in Fig.6 to account for drift, nonlinearities, and noise of the $z$ position during a $I(z)$ scan. This offset is evident at the largest currents where the data points (gray circles) should ideally approach zero.

------

\*       snijderspc@ornl.gov

\*\*      ingo.barke@uni-rostock.de